\documentclass[11pt]{article}
\usepackage{graphicx}

\title{\bf Loops in the Reggeon model for hA scattering}
\author{M. Braun, A. Tarasov\footnote{andrey.tarasov@cern.ch  }  \\ Department of High Energy physics, University of
S.Petersburg\\ 198904 Ulianovskaya 1,  S.Petersburg, Russia}
\date{}
\begin{document}

\maketitle
\input epsf
\def\beq{\begin{equation}}
\def\eeq{\end{equation}}
\def\phid{\phi^{\dagger}}
\def\pd{\partial}
\def\by{\bar{y}}
\def\tt{\tilde{T}}

{\bf Abstract}

Contribution of simplest loops is studied in the Local Reggeon
Field Theory  with a supercritical pomeron in the nuclear
matter. It is shown that the latter transforms the supercritical
pomeron into the subcritical. Renormalization of the
intercept and difficulties related to the fact that the renormalized
intercept is complex are discussed. Numerical results with the
conventional parameters are reported.

\section{Introduction}
With the advent of QCD much effort has been applied to study
high-energy hadron-nucleus scattering within its framework.
However QCD can reliably describe only the hard region of the
dynamics. Soft processes, contributing to the bulk of the total
cross-section, are more difficult to treat. The most sophisticated
approach is  to study  soft processes within the perturbative QCD
approach to small $x$ phenomena, based on the Balitsky-Kovchegov
equation, which sums all fan diagrams with self-interacting BFKL
pomerons ~\cite{bal,kov,bra1}. However this approach is based on
several approximations: a large number of colours $N_c$, fixed
small QCD coupling constant $\alpha_s$ and, most seriously,
neglect of loop diagrams. This latter approximation can be
justified if the parameter $\gamma=\alpha_s\exp{\Delta y}$ where
$y$ is the rapidity and $\Delta$ the pomeron intercept is small.
Then for a large nuclear target, such that $A^{1/3}\gamma\sim 1$,
the tree diagrams indeed give the dominant contribution and loops
may be dropped. However with the growth of $y$ the loop
contribution becomes not small and although tree diagrams are
still relatively enhanced by factor $A^{1/3}$, disregard of loops
cannot be rigorously justified.

Calculation of the loop contribution within the perturbative QCD appears to
be a formidable task. So it seems to be worthwhile to start with a simpler
approach, that of the old local Reggeon Field Theory (LRFT) with a
supercritical pomeron. Such a study, apart from its possible lessons for the
modern QCD approach, also has an independent value, since the
old LRFT with phenomenological parameters describes not so badly the
soft dynamics of high-energy strong interactions. In fact in many respects
it does it better that the perturbative QCD, which
encounters severe
difficulties due to a very large value of the BFKL intercept, hardly
compatible with the experimental data.
Also in LRFT applications  to hadron-nucleus ~\cite{schwim}
and nucleus-nucleus ~\cite{amati} collisions were restricted to the
approximation, in which
the contribution from loop diagrams is neglected.
If additionally the slope of the pomeron $\alpha'$ is
taken to be zero,
then the theory is effectively living in zero transverse dimensions
and allows for the explicit analytic solution. Still remaining in
zero transverse dimension the influence of loop diagrams has also been
studied, both   theoretically old ago in ~\cite{aless, jengo, ciabel} and
numerically recently in ~\cite{bravac}. This influence turns out to
be decisive for the asymptotic behaviour at large energies and transforms the
supercritical  pomeron into a weakly subcritical one with the effective
intercept $\propto -\exp(1/\lambda^2)$ where $\lambda$ is the
triple-pomeron coupling  assumed to be small.

Unfortunately generalization of these beautiful results to the
realistic case of two transverse dimensions is prohibitively
difficult. To start with, one is forced to introduce a non-zero
value of the slope: otherwise the loop contribution is divergent
in the impact parameter.  However even in the tree approximation
solution of the model with $\alpha'\neq 0$ is only possible
numerically. Second, the model in $d_T=2$ needs renormalization in
the ultraviolet. And, most important, the method used to solve the
model in $d_T=0$, which is to study the corresponding
quantum-mechanical system   and the relevant Schroedinger
equation, is inapplicable in the realistic case, since instead of
ordinary differential equations one arrives at  equations with
functional derivatives. In fact  summation of all loop
contributions is equivalent to  a complete solution of the
corresponding  quantum field theory, the task which seems to be
beyond our present possibilities. So at most one can hope to
obtain some partial results which might shed light onto the
properties of the model with loops. There were several attempts to
study the high-energy behaviour of the LRFT with a supercritical
pomeron using different approximate techniques and giving
contradicting results. In ~\cite{abarb} it was claimed that a
phase transition occurs at all values of the renormalized
intercept $\epsilon^{(r)}=\alpha^{(r)}(0)-1>0$ which  leads to a
theory which violates the projectile-target symmetry and so
physically unacceptable. On the other hand in ~\cite{ciaf} it was
found that the phase transition takes place only at values of
$\epsilon^{(r)}$ greater than some critical value
$\epsilon^{(r)}_c$. At $\epsilon^{(r)}<\epsilon^{(r)}_c$ the
theory essentially corresponds to that with a subcritical pomeron
and cross-sections vanishing at high energies. At
$\epsilon^{(r)}>\epsilon^{(r)}_c$ cross-sections grow as $\log^2
s$. No violation of the projectile-target was found. These results
were obtained by approximating the transverse plane as a
two-dimensional lattice and starting from the known asymptotical
solution at each lattice site without intersite interaction.
However in this approach transition to the continuous plane
requires knowing the single-site solutions at large values of the
triple pomeron interaction constant, which are not known.

In the present study we take a different approach. Instead of
trying to solve the model for the purely hadronic scattering we
consider the hadron-nucleus scattering and propagation of the
pomeron inside the heavy nucleus target. Moreover to avoid using
numerical solution of the tree diagrams contribution with
diffusion in the impact parameter, we concentrate on the case of a
constant nuclear density which allows to start with the known
analytical solutions. The point which is decisive for the
following derivation is that the nuclear surrounding  transforms
the pomeron from the supercritical one with intercept $\epsilon>0$
to a subcritical one with the intercept $-\epsilon$. Then Regge
cuts, corresponding to loop diagrams, start at branch points
located to the left of the pomeron pole and their contribution is
subdominant at high energies. As a result the theory acquires the
properties similar to the standard LRFT with a subcritical pomeron
and allows for application of the perturbation theory. For a
finite nucleus it leads to cross-sections with tend to a constant
value at large energies. For the physical nucleus with a long
distance tail it seems to lead to cross-sections growing as
$\log^2 s$ in agreement with ~\cite{ciaf}.

In fact we cannot prove all these
statements rigorously, since we are not able to sum all the loops
but only  a certain simplest  subset of them. However  the results we find
seem to be rather general.

The paper is organized as follows. In the next section we
introduce the model and remind some relevant known results  for
the zero-slope case $\alpha'=0$. In Section 3 we reformulate the
model to describe the loops in the nuclear surrounding. Then we
calculate the contribution from simplest loops and discuss their
renormalization in Section 4. Next in Sections 5 and 6 we sum
insertions of any number of simplest loops into the scattering
amplitude and Green function by solving the relevant Dyson
equations. Section 7 gives some numerical illustration of the loop
influence for the amplitude and Green function with physically
chosen parameters of the theory. Finally Section 8 draws some
conclusions.

\section{The model. Zero-slope results}

We study the LRFT model based on two pomeron fields
$\phi(y,b)$ and $\phid(y,b)$ depending on the rapidity $y$ and
impact parameter $b$, with a Lagrangian density
 \beq
L=\phid S\phi+\lambda\phid\phi(\phi+\phid)+g\rho\phi.
\label{lagden} \eeq Here  in the free part \beq
S=\frac{1}{2}(\overrightarrow{\partial}_y-\overleftarrow{\partial_y})
+\alpha'\nabla_b^2+\epsilon, \eeq where $\epsilon$ is the
intercept minus unity and $\alpha'$ is the slope. The  source term
describing interaction with the nuclear target is \beq
\rho=AT(b)\delta(y). \eeq It is assumed that for a supercritical
pomeron $\epsilon>0$ and $\lambda<0$. The action is \beq {\cal
A}=\int dy\int d^2bL(y,b) \eeq and the generating functional is
\beq Z=\int D\phi D\phid e^{\cal A}. \eeq The free Green function
in the momentum space is \beq
G^{(0)}(y,k)=\theta(y)e^{y(\epsilon-\alpha'k^2)}. \label{freeg}
\eeq

The classical equation of motion are \beq \frac{\delta
L}{\delta\phi}=-\partial_y\phid+\alpha'\nabla_b^2\phid
+\epsilon\phid+\lambda{\phid}^2+2\lambda\phi\phid +g\rho=0
\label{eqphidi} \eeq and \beq \frac{\delta
L}{\delta\phid}=\partial_y\phi+\alpha'\nabla_b^2\phi
+\epsilon\phi+\lambda\phi^2+2\lambda \phid\phi=0. \label{eqphi}
\eeq From the latter equation we find $\phi=0$ and the equation
for $\phid$ takes the form \beq
\partial_y\phid=\alpha'\nabla_b^2\phid
+\epsilon\phid+\lambda{\phid}^2,
\label{eqphid}
\eeq
with an initial condition
\beq
\phid(y=0)=gAT(b).
\label{iniphid}
\eeq
Equation (\ref{eqphid}) describes evolution of the pomeron field in
rapidity and its diffusion in the impact parameter inside the nucleus.
In the
approximation $\alpha'=0$
equation for $\phid$ simplifies to
\beq
\partial_y\phid=
\epsilon\phid+\lambda{\phid}^2,
\label{eqphid0}
\eeq
which is trivially solved for each given $b$ ~\cite{schwim}:
\beq
\phid(y,b)=\frac{gAT(b)e^{\epsilon y}}
{1-\lambda gAT(b)\frac{1}{\epsilon}\Big(e^{\epsilon y}-1\Big)}
\equiv \xi(y,b).
\label{phidcl}
\eeq
The scattering matrix is
\beq
T(y,b)=g\xi(y,b).
\eeq
\section{Loops in the nuclear surrounding}
\subsection{Transformation to the nuclear background}
In this section we shall analyze the structure of the model beyond
the tree approximation in the nuclear background. To locate loops
we make a shift in field $\phid$: \beq
\phid(y,b)=\phid_1(y,b)+\xi(y,b) \label{shift} \eeq and
reinterpret our theory in terms of fields $\phi$ and $\phid_1$.

The Lagrangian density becomes
\beq
L=(\phid_1+\xi)S\phi+\lambda (\phid_1+\xi)^2\phi+
\lambda\phi^2(\phid_1+\xi)+g\rho\phi.
\eeq
Terms linear in $\phi$
vanish due to equation of motion for $\xi$. We are left with
\beq
L=\phid_1(S+2\lambda \xi)\phi+\lambda\xi\phi^2+
\lambda\phid_1\phi(\phid_1+\phi).
\eeq
This Lagrangian corresponds to a theory in the vacuum with the
pomeron propagator  in the external field $f(b,y)=2\lambda\xi(y,b)$
\beq
G^{(0)}_f =-(S+2\lambda\xi)^{-1},
\eeq
 the standard triple interaction
and an extra interaction described by the term $\lambda\xi\phi^2$. This new
interaction corresponds to the transition of a pair of pomerons into the vacuum
at point $(y,b)$ with a vertex $\lambda\xi(y,b)$, see Fig. 1.

Loops in the Green function may be formed both by the standard
interaction and the new one. In the latter case they are to be
accompanied by at least a pair of of standard interactions.
Diagrams with a few  simple loops in the Green function are
illustrated in Fig. 2. One immediately observes that a loop formed
by the standard interaction has the order $\lambda^2/\alpha'$ and
requires renormalization. A loop formed by a new interaction has
the order $\lambda^3/\alpha'$ and is finite.

The amplitude is obtained as a tadpole $g<\phid_1(y,b)>$. The
simplest diagrams for it contains one loop and are shown in Fig. 3
$a,b$. Diagrams with more loops are shown in Figs. 3 $c,d$.

\subsection{The Green function in the external field $2\lambda\phid(y)$}
In the construction of loops there appears a Green function
$G^{(0)}_f(y,b|y'b')$ in which the pomeron fields are coupled to
the external field \beq f(y,b)=2\lambda \xi(y,b). \label{deff}
\eeq This Green function satisfies the equation \beq
\frac{dG^{(0)}_f(y,b|y'b')}{dy}= (\epsilon+\alpha'\nabla_b^2)
G^{(0)}_f(y,b|y',b')+f(y,b)G^{(0)}_f(y,b|y',b') \eeq with the
boundary conditions \beq G^{(0)}_f(y,b|y',b')=0,\ \ y-y'<0,\ \
G^{(0)}(y',b|y',b')=\delta^2(b-b'). \label{inigfb} \eeq

In the general case the Green function $G^{(0)}_f$ can only be calculated
numerically, just as the external field $f(y,b)$. Its analytic form
can be found in two cases.

If the slope $\alpha'=0$ then obviously we have
\beq
G^{(0)}_f(y,b|y',b')=\delta^2(b-b')G^{(0)}_f(y,y',b),
\eeq
where
$G^{(0)}_f(y,y',b)$ is the Green function in the zero-dimensional
world which satisfies
\beq
\frac{dG^{(0)}_f(y,y',b)}{dy}= \epsilon
G^{(0)}_f(y,y',b)+f(y,b)G^{(0)}_f(y,y',b) \label{eqg0}
\eeq
with
the boundary conditions
\beq
 G^{(0)}_f(y,y',b)=0,\ \ y-y'<0,\ \
G^{(0)}_f(y',y',b)=1 \eeq and $f(y,b)$ given by (\ref{deff}) and
(\ref{phidcl}). Solution of (\ref{eqg0}) for $G^{(0)}_f$ is
trivial. It is easy to find that \beq
G^{(0)}_f(y,y',b)=e^{\int_{y'}^y ds(\epsilon+f(s,b))}=
e^{-\epsilon (y-y')} \left(\frac{a-(a-1)e^{-\epsilon
y'}}{a-(a-1)e^{-\epsilon y}}\right)^2=
e^{\epsilon(y-y')}\frac{p^2(y')}{p^2(y)}, \label{gf01} \eeq where
\beq a=-\frac{\lambda gAT(b)}{\epsilon} >0 \label{adef} \eeq and
\beq p(y)=1+a\Big(e^{\epsilon y}-1\Big). \label{pdef} \eeq It is
remarkable that at large $y$ the Green function
$G^{(0)}_f(y,y',b)\simeq \exp(-\epsilon y)$, that is behaves as
the free Green function with the opposite sign of $\epsilon$ and
so vanishes at $y\to\infty$. In contrast to the free Green
function, it corresponds to a subcritical pomeron.

The second case which admits an analytic solution for the Green
function is that of the nuclear matter, that is the case when the
profile function is constant in all transverse space \beq
T(b)=T_0. \eeq Physically this case corresponds to the behaviour
of the Green function at values of of the impact parameter well
inside the nucleus, where the variation of $T(b)$ is small. Then
the classical field $\xi$ and the external field $f=2\lambda\xi$
become $b$-independent and the equation for the Green function in
this field takes the form \beq \frac{dG^{(0)}_f(y,b|y',b')}{dy}=
(\epsilon+\alpha'\nabla_b^2)G^{(0)}_f(y,b|y',b')+f(y)G^{(0)}_f(y,b|y',b')
\label{eqgfb1} \eeq with the same initial condition
(\ref{inigfb}). This equation can also be easily solved. We
present \beq G^{(0)}_f(y,b|y',b')=X(y-y',b,b')e^{\int_{y'}^y
dy_1(\epsilon+f(y_1))}= X(y-y',b,b')G^{(0)}_f(y,y',b), \eeq where
$G^{(0)}_f(y,y',b)$ is the Green function in zero-dimensional
world (\ref{gf01}) satisfying $G^{(0)}_f(y',y',b)=1$ Then we
obtain an equation for $X$: \beq
\partial_y X(y,b,b')=\alpha'\nabla_b^2X_f(y,b,b')
\eeq with an initial condition \beq X(0,b,b')=\delta^2(b-b') \eeq
and thus with the solution \beq X(y-y',b,b')=\frac{1}{4\pi\alpha'
(y-y')}e^{-(b-b')^2/(4\alpha' (y-y'))}. \eeq So in the nuclear
matter the Green function in the external field takes a simple
form \beq G^{(0)}_f(y,b|y',b')=\frac{1}{4\pi\alpha' (y-y')}
e^{-\epsilon (y-y')-(b-b')^2/(4\alpha'(y-y'))}
\left(\frac{a-(a-1)e^{-\epsilon y'}} {a-(a-1)e^{-\epsilon
y}}\right)^2, \eeq where we used  (\ref{gf01}). In the momentum
space we find \beq
G^{(0)}_f(y,y',k)=e^{-(y-y')(\epsilon+\alpha'k^2)}
\left(\frac{a-(a-1)e^{-\epsilon y'}} {a-(a-1)e^{-\epsilon
y}}\right)^2 =e^{(y-y')(\epsilon-\alpha'
k^2)}\frac{p^2(y')}{p^2(y)}. \eeq

\begin{figure}
\hspace*{4 cm} \includegraphics[width=3 cm]{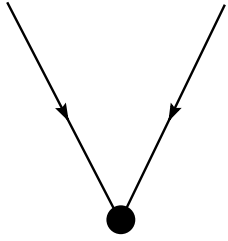} \caption{The
new vertex for two-pomeron annihilation, which appears after the
shift in field $\phid$} \label{fig1}
\end{figure}

Note an especially simple case when $a=1$ and the Green function
in the external  field coincides with the free Green function with
the opposite sign of $\epsilon$: \beq G^{(0)}_f(y,y',k)\Big|_{a=1}
=e^{-(y-y')(\epsilon+\alpha'k^2)}. \eeq Then the theory formally
corresponds to a subcritical pomeron model with an additional
interaction shown in Fig. \ref{fig1}


\section{Lowest order loops}
\subsection{The lowest order loop in absence of the nucleus and its
renormalization}
Let us study the simplest contribution from loops, the single loop in
the Green function.
In the second order in $\lambda$ we get for the Green function
\beq
G^{(2)}(y,k)=-2\lambda^2\int dy_1dy_2\frac{d^2k_1}{(2\pi)^2}
G^{(0)}(y-y_1,k)G^{(0)}(y_1-y_2,k_1)G^{(0)}(y_1-y_2,k-k_1)G^{(0)}(y_2,k).
\label{gf2}
\eeq
The minus sign comes from the fact that the simple Green function is in
fact $-G^{(0)}$, so that
in the second order we get $-G^{(0)}*(-{G^{(0)}}^2)(-G^{(0)})=-G^{(2)}$. Factor 2 comes from
the contraction of ${\phid}^2(y_1,b_1)\phi^2(y_2,b_2)$.
Integration over $k_1$ is done trivially
\beq
\Sigma(y,k)\equiv -2\lambda^2\int\frac{d^2k_1}{(2\pi)^2}
G^{(0)}(y,k_1)G^{(0)}(y,k-k_1)=
-C\frac{1}{ y}e^{2\epsilon y-(1/2)\alpha'y k^2},
\eeq
where
\beq
C=\frac{\lambda^2}{4\pi\alpha'}.
\eeq
$\Sigma(y,k)$ is finite as it stands but leads to divergence at $y=0$
when substituted
into (\ref{gf2}). So it requires renormalization, which is achieved by
introducing
an extra mass term $\Delta{\epsilon}\phid\phi$ into the Lagrangian.
The renormalized self-mass is therefore
\beq
\Sigma^{(r)}(y,k)=\Sigma(y,k)+\Delta\epsilon\delta(y)
\eeq
and we have to require that
\beq
\int_0^ydy'\Sigma^{(r)}(y',k)={\rm finite\ terms}
-C\ln \frac{1}{y_{min}}+\Delta\epsilon<\infty,
\eeq
So we conclude that
\beq
\Delta\epsilon=
-C\ln(c_Ry_{min}),
\label{deltaep}
\eeq
where $y_{min}$ is the cutoff at small values of $y$ and $c_R$ is
an arbitrary finite constant.
Thus we find that the renormalized self mass is
\beq
\Sigma^{(r)}(y,k)=-C
\Big(\frac{1}{y}e^{2\epsilon y-(1/2)\alpha'y k^2}+
\delta(y)\ln(c_Ry_{min})\Big).
\label{freesig}
\eeq
or, in the impact parameter space,
\beq
\Sigma^{(r)}(y,b)=-C
\Big(\frac{1}{2\pi\alpha'y^2}e^{2\epsilon y-b^2/(2\alpha' y)}+
\delta(y)\delta^2(b)
\ln(c_Ry_{min})\Big).
\eeq

The Dyson equation for the Green function becomes \beq
G(y,0,k)=G^{(0)}(y,0,k)+\int_0^ydy_1G^{(0)}(y-y_1,k)\int
_0^{y_1}dy_2 \Sigma^{(r)}(y_1-y_2,k)G(y_2,k). \label{freedys} \eeq
It is trivially solved by the Laplace transformation. We introduce
Laplace transforms \beq G(E,k)=\int_0^{\infty}dye^{-Ey}G(y,0,k)
\eeq and similarly $G^{(0)}(E,k)$ and $\Sigma^{(r)}(E,k)$. In
terms of the Laplace transforms \beq
G(E,k)=\frac{1}{\Big(G^{(0)}(E,k)\Big)^{-1}-\Sigma^{(r)}(E,k)}
\eeq

From (\ref{freeg}) and (\ref{freesig}) we have
\beq
G^{(0)}(E,k)=\frac{1}{E-\epsilon+\alpha'k^2}.
\label{freege}
\eeq
and
\[
\Sigma^{(r)}(E,k)=-C\Big(\int_{y_{min}}^{\infty}\frac{dy}{y}
e^{-y(E-2\epsilon+\alpha'k^2/2)} +\ln (c_Ry_{min})\Big)\]\beq =
C(\ln(E-2\epsilon+\alpha'k^2/2)+C_E-\ln c_R). \label{freese} \eeq
(Note that terms with $\ln (y_{min})$ cancel). So the Laplace
transform for the Green function is \beq
G^{-1}(E,k)=E-\epsilon+\alpha'k^2-
C[\ln(E-2\epsilon+\alpha'k^2/2)+C_E-\ln c_R]. \label{gvac} \eeq
Its singularities in the complex $E$-plane consist of a
branchpoint at $E=2\epsilon-\alpha'k^2/2$ and a  pole
 $E_0(k)$ determined by the equation:
\beq G^{-1}(E_0,k)=0 \label{greenpo} \eeq At small $\lambda$ Eq.
(\ref{greenpo}) gives two complex conjugate poles \beq
E_0^{(\pm)}=\epsilon-\alpha'k^2
+C(\ln(|\epsilon+\alpha'k^2/2|)+C_E-\ln c_R\pm i\pi).
\label{greenpole} \eeq These poles  appear on the physical sheet
due to the abnormal sign of the self-mass contribution, which is a
consequence of the imaginary coupling in the original Lagrangian
of the LRFT. The asymptotic of the Green function is $\sim
\exp(2\mu y)$ in accordance of the contribution of the Regge cut
for a supercritical pomeron. According to  (\ref{deltaep}) the
unrenormalized pomeron intercept (depending on the cutoff) is \beq
\epsilon_0(y_0)=\epsilon-C\ln (c_Ry_{min}). \eeq It goes to
infinity as $y_{min}\to 0$, since $C>0$. As to the renormalized
intercept its definition may be chosen in different ways.
Traditionally it may be taken as the position of the pole in $E$
in the Green function at $k=0$.  Then it is  given by
(\ref{greenpo}), complex and $c_R$-dependent. One observes the
difficulty in doing renormalization in the standard way by
requiring the zero order propagator to have the pole at a chosen
energy $E_0(k=0)$: unless $E_0>2\epsilon$ there are two complex
conjugate values of $E_0$ and if one chooses $E_0>2\epsilon$
contributions from higher order self mass will again make it
complex. So the structure of perturbative singularities prohibits
choosing the zero-order propagator to carry the pole singularity,
since there are two. However from the physical point of view such
a choice is not necessary Inclusion of more pomeron exchanges
shifts  the dominating singularity further to the right, so that
the simple pole contribution to any physical process becomes
completely obliterated.  It is more reasonable to relate the value
of the renormalized intercept (and thus fix the arbitrary constant
$c_R$) to some physical observables. In the following we shall see
that in the scattering on the nucleus there appears a possibility
to define the renormalized intercept from experimental data at
high energies, with a certain choice of $c_R$.


\subsection{Lowest order loop in the nuclear matter}
The contribution from the lowest order loop (Fig. 2 $a$)
in the nuclear matter to the Green function in is
\beq
G^{(2)}_f(y,0,k)=\int dy_1dy_2G^{(0)}_f(y,y_1,k)
\Sigma_1(y_1,y_2,k)G^{(0)}_f(y_2,0,k).
\label{app31}
\eeq
where the Green functions $G^{(0)}_f$ are given by (\ref{gf01}) and
\beq
\Sigma_1(y_1,y_2,k)=-C\Big(
\frac{1}{y_1-y_2}e^{-2\epsilon (y_1-y_2)-(1/2)(y_1-y_2)\alpha'k^2}
e^{2\int_{y_2}^{y_1}dsf(s)}+\delta(y_1-y_2)\ln(c_Ry_{min})\Big).
\label{app32}
\eeq
is the renormalized 2nd order self mass in the nuclear background.
The subtraction term is easily calculated: \beq
G^{(2)}_{f,subtr}=-C\ln(c_Ry_{min}) \int
dy_1G^{(0)}_f(y,y_1,k)G^{(0)}_f(y_1,0,k)=
-C\ln(c_Ry_{min})yG^{(0)}_f(y,0,k). \eeq
The main term is given by \beq
-CG^{(0)}_f(y,0,k)\int_0^ydy_1\int_0^{y_1}dy_2\frac{1}{y_1-y_2}
e^{(y_1-y_2)(\epsilon+\frac{1}{2}\alpha'k^2)}\frac{p^2(y_2)}{p^2(y_1)}.
\eeq Changing to integration variable $z=y_1-y_2$ we find \beq
G_{f,main}^{(2)}(y,0,k)=-CG^{(0)}_f(y,0,k)\int_0^y\frac{dz}{z}W(z),
\eeq where \beq W(z)=e^{z(\epsilon+\frac{1}{2}\alpha'k^2)}\int
_z^ydy_1\frac{p^2(y_1-z)}{p^2(y_1)} \equiv
e^{\frac{1}{2}z\alpha'k^2}\tilde{W}(z). \label{tilw} \eeq We note
that $W(0)=y$ so that we can write \beq
G_{f,main}^{(2)}(y,0,k)=-CG^{(0)}_f(y,0,k)\Big\{\int_0^y\frac{dz}{z}\Big(W(z)-y\Big)
+y\ln\frac{y}{y_{min}}\Big\}. \eeq The second term just
substitutes in the subtraction contribution $y_{min}$ by $y$. The
first term can be integrated numerically since it converges at
$z=0$.

Function $\tilde{W}(z)$ can be calculated analytically.
Passing to integration
variable $u=e^{\epsilon y_1}$ and doing simple integrals we find
\beq
\tilde{W}(z)=e^{-\epsilon z}\Big(y-z+\frac{q_1^2-q^2}{\epsilon q^2}
\ln\frac{u_2(u_1+q)}{u_1(u_2+q)}
-\frac{(q_1-q)^2}{\epsilon q}\frac{u_2-u_1}{(u_2+q)(u_1+q)}\Big).
\eeq
where $u_1=e^{\epsilon z}$, $u_2=e^{\epsilon y}$, $q=1/a-1$
and $q_1=qu_1$.

Note that after renormalization the 2nd order contribution to the
Green function is also finite in the nuclear background.

\subsection{3d order loop Fig. 2 $b$}
The new  loop shown in Fig. 2 $b$, which appears in the nucleus,
has formally order $\lambda ^3$. However it is proportional to $AT(b)$,
so that  the extra $\lambda$ combines into the characteristic parameter
$a$, which is not small. So in fact it has the same order of magnitude
as the  simple loop of Fig. 2 $a$.

The contribution to the pomeron self-mass corresponding to Fig. 2 $b$
is
\beq
\Sigma_2(y_2,y_3,k)=4\lambda^3\int_0^{y_3}dy_1\int \frac{d^2k_1}{(2\pi)^2}
G_f^{(0)}(y_2,y_3,k-k_1)G_f^{(0)}(y_2,y_1,k_1)G_f^{(0)}(y_3,y_1,k_1)\xi(y_1).
\label{sig1}
\eeq
Here it is assumed that $y_2>y_3$. The final factor 4 combines
factor 4 from contractions
of $\phi^2(y_1)$ with ${\phid}^2(y_2)$ and ${\phid}^2(y_3)$, factor 2 coming
from interchange $2\leftrightarrow 3$ and factor 1/2 left from $1/3!$
in the development of action in the third order in interaction.
The sign "+" comes from the odd number of propagators in the contribution
to the full propagator with a minus sign. Using explicit expressions for
the propagators and $\xi$ and definition (\ref{adef}) we rewrite
(\ref{sig1}) as
\beq
\Sigma_2(y_2,y_3,k)=-4\lambda^2a\epsilon
\int_0^{y_3}dy_1\frac{p^3(y_1)}{p^4(y_2)}e^{\epsilon(2y_2-y_1)
-(y_2-y_3)\alpha'k^2}
\int \frac{d^2k_1}{(2\pi)^2}
e^{-2\alpha'k_1^2(y_2-y_1)+2(y_2-y_3)\alpha'kk_1}.
\label{sig11}
\eeq
The integral over $k_1$ is
\[
\frac{1}{8\pi\alpha'(y_2-y_1)}e^{\alpha'k^2\frac{(y_2-y_3)^2}{2(y_2-y_1)}},
\]
so that in the end we obtain
\beq
\Sigma_2(y_2,y_3,k)=-\frac{\lambda^2a\epsilon}{2\pi\alpha'}
\frac{1}{p^4(y_2)}e^{2\epsilon y_2-(y_2-y_3)\alpha'k^2}
\int_0^{y_3}\frac{dy_1}{y_2-y_1}p^3(y_1)
e^{-\epsilon y_1+\alpha'k^2\frac{(y_2-y_3)^2}{2(y_2-y_1)}}
\label{sig12}
\eeq

It is trivial to see that   $\Sigma_2$ is finite and does not need
any more renormalization.

\section{Amplitude in the nuclear matter approximation}
\subsection{Lowest order}
We start from the lowest order approximation, when
at fixed $b$ the scattering amplitude is
\beq
T^{(0)}(y,b)=g^2AT(b)\frac{e^{\epsilon y}}{p(y)}.
\eeq
The total forward scattering amplitude is obtained after integration over all $b$:
\beq
T^{(0)}(y)=\int d^2bT^{(0)}(y,b).
\label{totam}
\eeq
This expression can be rewritten in the momentum space, once we
introduce the Fourier transform of $T(y,b)$ by
\beq
T^{(0)}(y,b)=\int \frac{d^2k}{(2\pi)^2}e^{ikb}\tilde{T}^{(0)}(y,k),
\eeq
as
\beq
T^{(0)}(y)=\int\frac{d^2bd^2k}{(2\pi)^2}e^{ikb}{\tt}^{(0)}(y,k)={\tt}^{(0)}(y,0).
\label{tt}
\eeq

In the nuclear matter $T(y,b)$ does not depend on $b$:
\beq
T^{(0)}(y,b)=g^2AT_0\frac{e^{\epsilon y}}{p(y)},
\label{deft}
\eeq
so that its Fourier transform is
\beq
\tt^{(0)}(y,k)=(2\pi)^2\delta^2(k)g^2AT_0\frac{e^{\epsilon y}}{p(y)}.
\eeq
From (\ref{totam}) the forward scattering amplitude is
\beq
T^{(0)}(y)=\int d^2b t(y)=\pi R_A^2 g^2AT_0\frac{e^{\epsilon y}}{p(y)}=
g^2A\frac{e^{\epsilon y}}{p(y)},
\eeq
where we used that $T_0=1/(\pi R_A^2)$.
Due to (\ref{tt}) this of course implies that
$
(2\pi)^2\delta^2(k=0)\to \pi R_A^2.
$
In the limit $y\to\infty$ the lowest order amplitude tends to a finite value
\beq
T^{(0)}(y)_{y\to\infty}=\frac{g^2A}{a}=
\pi R_A^2\frac{g\epsilon}{|\lambda|}.
\eeq

\subsection{2nd order}
In the next order we have
\[
\tt^{(2)}(y,k)=\int dy_1dy_2G_f^{(0)}(y,y_1,k)\Sigma_1(y_1,y_2,k)\tt^{(0)}(y_2,k)
\]\beq
=(2\pi)^2\delta^2(k)g^2AT_0\int
dy_1dy_2G_f^{(0)}(y,y_1,0)\Sigma_1(y_1,y_2,0) \frac{e^{\epsilon
y_2}}{p(y_2)} \eeq so that the forward scattering amplitude is
\beq T^{(2)}(y)=\pi R_A^2 \int
dy_1dy_2G_f(y,y_1,0)\Sigma_1(y_1,y_2,0)T^{(0)}(y_2, b). \eeq

The subtraction term in $\Sigma^{(r)}$ gives a contribution
 \beq
T^{(2)}_{sub}(y)=-C\pi R_A^2 \ln(c_Ry_{min})\int_0^y dy_1
G_f^{(0)}(y,y_1,0)T^{(0)}(y_1, b). \eeq Using (\ref{deft}) and
(\ref{gf01}) we get \beq T^{(2)}_{sub}(y)= -C
g^2AG_f^{(0)}(y,0,0)\ln(c_Ry_{min}) \int_0^ydy_1p(y_1). \eeq We
define \beq I(y,z)\equiv\int_z^ydy_1p(y_1)=
\frac{a}{\epsilon}\Big(e^{\epsilon y}-e^{\epsilon z}\Big)+
(y-z)(1-a). \eeq Then we find \beq
T^{(2)}_{sub}(y)=-Cg^2A\frac{e^{\epsilon y}}{p^2(y)}
\ln(c_Ry_{min})I(y,0). \eeq

The main term is
\beq
T^{(2)}_{main}(y)=
-Cg^2A\frac{e^{\epsilon y}}{p^2(y)}
\int_0^ydy_1\int_0^{y_1} dy_2e^{\epsilon(y_1-y_2)}\frac{1}{y_1-y_2}
\frac{p^3(y_2)}{p^2(y_1)}.
\eeq
Passing to integration variables $y_1$ and $z=y_1-y_2$ we find
\beq
T^{(2)}_{main}(y)=-Cg^2A\frac{e^{\epsilon y}}{p^2(y)}\int_0^y\frac{dz}{z}\Psi(z),
\eeq
where
\beq
\Psi(z)=e^{\epsilon z}\int_z^ydy_1\frac{p^3(y_1-z)}{p^2(y_1)}.
\eeq
Note that
\beq
\Psi(0)=I(y,0).
\eeq
Thus we find
\beq
T^{(2)}_{main}(y)=-Cg^2A
\frac{e^{\epsilon y}}{p^2(y)}\Big\{\int_0^y\frac{dz}{z}\Big(\Psi(z)-\Psi(0)\Big)
+\ln\frac{y}{y_{min}}I(y,0)\Big\}.
\eeq
The second term just changes $y_{min}$ to $y$ in the subtraction term.

Function $\Psi(z)$ can be calculated analytically in the same way as
$\tilde{W}(z)$ (see Eq. (\ref{tilw})).
One finds
\[
\Psi(z)=\frac{I(y,z)}{u_1^2}+3aq\frac{u_1-1}{u_1^2}(y-z)
+aq(2+u_1)\frac{(u_1-1)^2}{\epsilon u_1^2}
\ln\frac{u_2(u_1+q)}{u_1(u_2+q)}
\]\beq
+aq^2\frac{(u_1-1)^3}{\epsilon u_1^2}
\frac{u_1-u_2}{(u_2+q)(u_1+q)}.
\eeq
where as before $u_1=e^{\epsilon z}$, $u_2=e^{\epsilon y}$, $q=1/a-1$
and $q_1=qu_1$.
The remaining integration over $z$ has to be done numerically.

It is not difficult to find the asymptotic behaviour of
$T^{(2)}(y)$ at large $y$.  At $y\to\infty$, asymptotically \beq
p(y)=ae^{\epsilon y},\ \ I(y, 0)=\frac{a}{\epsilon}e^{\epsilon y}
\eeq so that  in this limit \beq
T^{(2)}(y)_{sub}=-C\frac{g^2A}{a\epsilon}\ln (c_Ry_{min})
\label{asysub} \eeq

The asymptotic of the main term comes from large values of $z$
inside the integral. At large $z$ \beq
\Psi(z)-\Psi(0)=-\frac{a}{\epsilon}e^{\epsilon
y}\Big(1-e^{-2\epsilon z}\Big) \eeq and so \beq
\int^y\frac{dz}{z}\Big(\Psi(z)-\Psi(0)\Big) \simeq
\frac{a}{\epsilon}e^{\epsilon y}\Big({\rm Ei}(-2\epsilon y)-\ln
(2\epsilon y)-C_E\Big) \simeq -\frac{a}{\epsilon}e^{\epsilon
y}\Big(\ln (2\epsilon y)+C_E\Big). \eeq Thus  asymptotically \beq
T^{(2)}_{main}(y)=C\frac{g^2A}{a\epsilon} \Big(\ln (2\epsilon
y_{min})+C_E\Big). \label{asimain} \eeq

In the sum the terms with $\ln y_{min}$ cancel and we find that in
the limit $y\to \infty$ the loop contribution also tends to a
constant \beq T^{(2)}(y)_{y\to\infty}=-C\frac{g^2A}{a\epsilon}
\Big(\ln \frac{c_R}{2\epsilon}-C_E\Big). \eeq The ratio first to
second order is at $y\to\infty$ \beq
r^{(2)}(y)=\frac{T^{(2)}(y)}{T^{(0)}(y)}\Big|_{y>>1}=
-\frac{\lambda^2}{4\pi\alpha' \epsilon}\Big(\ln
\frac{c_R}{2\epsilon}-C_E\Big). \eeq Remarkably it does not depend
on $A$.

\subsection{3d order}
In the third order the forward scattering amplitude is \beq
T^{(3)}(y)=\pi R_A^2 \int
dy_2dy_3G_f(y,y_2,0)\Sigma_2(y_2,y_3,0)T^{(0)}(y_3, b). \eeq Using
(\ref{sig12}) at $k=0$ and interchanging the order of integration
in $y_2$ and internal integration in $y_1$ inside $\Sigma_2$ we
find \beq T^{(3)}(y)=-g^2A\frac{\lambda^2a\epsilon}{2\pi\alpha'}
\frac{e^{\epsilon y}}{p^2(y)}\int_0^y dy_2\frac{e^{\epsilon
y_2}}{p^2(y_2)}
\int_0^{y_2}\frac{dy_1}{y_2-y_1}p^3(y_1)e^{-\epsilon y_1}
\int_{y_1}^{y_2} dy_3\frac{e^{\epsilon y_3}}{p(y_3)}. \label{t32}
\eeq

It is not difficult to find the asymptotic of $T^{(3)}(y)$ at large $y$.
We present (\ref{t32}) in the form
\beq
T^{(3)}(y)=-g^2A\frac{\lambda^2a\epsilon}{2\pi\alpha'}
\frac{e^{\epsilon y}}{p^2(y)}\int_0^y dy_2\frac{e^{\epsilon y_2}}{p^2(y_2)}
F(y_2),
\label{t33}
\eeq
where taking $y_1=y_2\beta_1$ and $y_3=y_2\beta_3$
\beq
F(y_2)=y_2
\int_0^{1}\frac{d\beta_1}{1-\beta_1}p^3(y_2\beta_1)e^{-\epsilon y_2\beta_1}
\int_{\beta_1}^1 d\beta_3\frac{e^{\epsilon y_2\beta_3}}{p(y_2\beta_3)}.
\label{t34}
\eeq
We are interested in the behaviour of $F(y_2)$ as $y_2>>1$, since this governs
the behaviour of the integral over $y_2$ in (\ref{t33}).
As $y_2\to\infty$ we have $p(y_2\beta)\to a\exp(\epsilon y_2\beta)$ so that
\beq
F(y_2)_{y_2>>1}=a^2 y_2
\int_0^{1}\frac{d\beta_1}{1-\beta_1}e^{2\epsilon y_2\beta_1}
\int_{\beta_1}^1 d\beta_3=
\frac{a^2}{2\epsilon}\Big(e^{2\epsilon y_2}-1\Big).
\label{t35}
\eeq
Putting this asymptotic in (\ref{t33}) we obtain
\beq
T^{(3)}(y)_{y>>1}=-g^2A\frac{\lambda^2}{4\pi a\alpha'}
e^{-\epsilon y}\int_0^y dy_2e^{\epsilon y_2}
\simeq
-g^2A\frac{\lambda^2}{4\pi a \alpha'\epsilon}.
\eeq
Recalling that at $y>>1$ $T^{(0)}=g^2A/a$ we find  the ratio
\beq
r^{(3)}(y)=\frac{T^{(3)}(y)}{T^{(0)}(y)}\Big|_{y>>1}=
-\frac{\lambda^2}{4\pi \alpha'\epsilon}
\eeq

Note that in the particular case $a=1$, $T^{(3)}$ can easily be
found explicitly at all $y$. In fact trivial integrations give
\beq T^{(3)}_{a=1}(y)= -g^2A\frac{\lambda^2}{4\pi
\alpha'\epsilon}\Big(1-e^{-\epsilon y}\Big)^2. \label{t36} \eeq

The full ratio $T/T^{(0)}$ from both loops, Fig. 2 $a$ and $b$, turns out to be
\beq
r(y)=r^{(2)}(y)+r^{(3)}(y)=-\frac{\lambda^2}{4\pi \alpha'\epsilon}
\Big(\ln\frac{c_R}{2\epsilon}+1-C_E\Big).
\eeq
The bracket is universal. So one may
{\it define} the renormalized self-mass by requiring that this bracket is zero,
which implies that the loop correction vanishes at large $y$:
\beq
\ln \frac{c_R}{2\epsilon}-C_E+1=0,\ \ {\rm or}\ \ c_R=2\epsilon e^{C_E-1}.
\label{crsel}
\eeq
This allows to experimentally determine $a$ and hence $\epsilon$ from the
behaviour of the amplitude at large $y$. In this way we may define
the renormalized intercept in a physically reasonable manner.

Note that with this choice of $c_R$ the renormalized intercept
$\epsilon^{(r)}$ formally defined as the position of the pole of
the propagator in the vacuum at $k=0$ is given by the solution of
the equation (\ref{greenpo}) which reads \beq
\epsilon^{(r)}-\epsilon-C[\ln\frac{\epsilon^{(r)}-2\epsilon}{2\epsilon}+1]=0
\label{reneps} \eeq At small $\lambda$ it is complex and close to
$\epsilon$.

\subsection{Random phase approximation}

One may try to sum all primitive loop insertions into the
amplitude ('random phase approximation'). In this approximation
the amplitude $T(y)$ in the nuclear matter satisfies the Dyson
equation \beq T(y)=T^{(0)}(y)+ \int_0^y
dy_1\int_0^{y_1}dy_2G_f^{(0)}(y,y_1,0)\Big(\Sigma_1(y_1,y_2,0)+
\Sigma_2(y_1,y_2,0)\Big)T(y_2), \label{trand} \eeq where
$\Sigma_1$ and $\Sigma_2$ are the 2nd and 3d order self-mass
contributions studied above.

We present
\beq
T(y)=T^{(0)}(y)r(y).
\eeq
The equation for $r(y)$ reads
\beq
r(y)=1+X_1(y)+X_2(y),
\eeq
where $X_1$ and $X_2$ are parts coming from $\Sigma_1$ and
$\Sigma_2$ respectively.

We find \beq X_1(y)=-\frac{C}{p(y)}\ln(c_Ry_{min})
\int_0^ydy_1p(y_1)r(y_1)-
\frac{C}{p(y)}\int_0^ydy_1\int_0^{y_1}dy_2\frac{1}{y_1-y_2}
e^{\epsilon(y_1-y_2)}\frac{p^3(y_2)}{p^2(y_1)}r(y_2).
\label{amprand} \eeq Again we pass to variable $z=y_1-y_2$ to
rewrite the last term as \beq
-\frac{C}{p(y)}\int_0^y\frac{dz}{z}\omega(y,z)=
-\frac{C}{p(y)}\int_0^y\frac{dz}{z}\Big(\omega(y,z)-\omega(y,0)\Big)
-\frac{C}{p(y)}\omega(y,0)\int_{y_{min}}^y\frac{dz}{z},
\label{lrterm} \eeq where \beq \omega(y,z)=e^{\epsilon
z}\int_z^ydy_1\frac{p^3(y_1-z)}{p^2(y_1)}r(y_1-z). \label{lomega}
\eeq Obviously \beq \omega(y,0)=\int_0^ydy_1p(y_1)r(y_1), \eeq so
that the last term in (\ref{lrterm}) changes $y_{min}$ in
(\ref{amprand}) to $y$. As a result we find that the part $X_1(y)$
in the equation for $r$ takes the form \beq
X_1=-\frac{C}{p(y)}\ln(c_Ry)
\int_0^ydy_1p(y_1)r(y_1)-\frac{C}{p(y)}\int_0^y\frac{dz}{z}
\Big(\omega(y,z)-\omega(y,0)\Big). \label{amprand1} \eeq

The part $X_2(y)$ is
\beq
X_2(y)=-\frac{2a\epsilon C}{p(y)}\int_0^ydy_1\frac{e^{\epsilon y_1}}{p^2(y_1)}
\int_0^{y_1}\frac{dy_3}{y_1-y_3}p^3(y_3)e^{-\epsilon y_3}
\int_{y_3}^{y_1}dy_2\frac{e^{\epsilon y_2}}{p(y_2)}r(y_2).
\label{amprand2}
\eeq

In the case $a=1$ and neglecting $X_2$
we can solve the Dyson equation passing to the Laplace transforms
in rapidity. We find
\beq
T(E)=\frac{T^{(0)}(E)\Big(G^{(0)}_f\Big)^{-1}(E,0)}
{\Big(G^{(0)}_f\Big)^{-1}(E,0)-\Sigma_f^{(r)}(E,0)},
\eeq
where $G^{(0)}_f(E,0)$ and $\Sigma_f^{(r)}(E,0)$ have the same form as
(\ref{freege}) and (\ref{freese}) with the opposite sign of $\epsilon$
and
\beq
T^{(0)}(E)=g^2A\frac{1}{E}.
\eeq

So the Laplace transform
 for the amplitude is
\beq
T(E)=\frac{g^2A}{E}\, \frac{E+\epsilon}
{E+\epsilon-C[\ln(E+2\epsilon)+C_E-\ln c_R]}.
\eeq
From the expression for $T(E)$ we immediately conclude that in the limit $y\to\infty$
the amplitude tends to
\beq
T(y)|_{y\to\infty}=g^2AZ,
\eeq
where
\beq
Z=\frac{\epsilon}
{\epsilon-C[\ln(2\epsilon)+C_E-\ln c_R]}.
\eeq
Factor $Z-1$ gives the correction from  loops. We can define the renormalization
condition by requiring that $Z=1$, which implies for this case ($X_2=0$)
\beq
\ln(2\epsilon)+C_E-\ln c_R=0,
\label{cr0}
\eeq
With this choice
\beq
\Sigma^{(r)}(E,k)=C\ln\frac{E+2\epsilon+\alpha'k^2/2}{2\epsilon}
\eeq
and
\beq
T(E)=\frac{g^2A}{E}\, \frac{E+\epsilon}
{E+\epsilon-C\ln\frac{E+2\epsilon}{2\epsilon}}.
\eeq

\section{Pomeron Green function in the random phase approximation}
In this approximation the full Green function in the nuclear
matter $G_f(y,0,k)$ satisfies the Dyson equation \beq
G_f(y,0,k)=G^{(0)}_f(y,0,k)+\int_0^y dy_1\int_0^{y_1}dy_2
G^{(0)}_f(y,y_1,k)\Big(\Sigma_1(y_1,y_2,k)+\Sigma_2(y_1,y_2,k)\Big)
G_f(y_2,0,k) \label{grand} \eeq We present \beq
G_f(y,0,k)=G^{(0)}_f(y,0,k)R(y,k) \eeq to obtain an equation for
$R$ \beq R(y,k)=1+Y_1+Y_2. \eeq Here $Y_1$ comes from $\Sigma_1$:
\beq Y_1(y,k)=-C\ln(c_Ry_{min})
\int_0^ydy_1R(y_1,k)-C\int_0^ydy_1\int_0^{y_1}dy_2\frac{1}{y_1-y_2}
e^{\beta(y_1-y_2)}\frac{p^2(y_2)}{p^2(y_1)}R(y_2,k) \label{rrand}
\eeq and $\beta=\epsilon+\alpha' k^2/2$. Passing to integration
variable $z=y_1-y_2$ we rewrite the last term in Eq. (\ref{rrand})
as \beq
-C\int_0^y\frac{dz}{z}\Omega(y,z)=-C\int_0^y\frac{dz}{z}\Big(\Omega(y,z)-\Omega(y,0)\Big)
-C\Omega(y,0)\int_{y_{min}}^y\frac{dz}{z}, \label{llterm} \eeq
where \beq \Omega(y,z)=e^{\beta
z}\int_z^ydy_1\frac{p^2(y_1-z)}{p^2(y_1)}R(y_1-z,k). \label{omega}
\eeq Obviously \beq \Omega(y,0)=\int_0^ydy_1R(y_1,k), \eeq so that
the last term in (\ref{llterm}) changes $y_{min}$ in (\ref{rrand})
to $y$. As a result we find \beq Y_1(y,k)=-C\ln(c_Ry)
\int_0^ydy_1R(y_1,k)-C\int_0^y\frac{dz}{z}
\Big(\Omega(y,z)-\Omega(y,0)\Big). \label{rrand1} \eeq

The part $Y_2$ comes from $\Sigma_2$:
\[
Y_2(y,k)=\]\beq
-2a\epsilon C\int_0^ydy_1\frac{e^{\epsilon y_1}}{p^2(y_1)}
\int_0^{y_1} \frac{dy_2}{y_1-y_2}p^3(y_2)e^{-\epsilon y_2}
\int_{y_2}^{y_1}dy_3\frac{e^{\epsilon y_3}}{p^2(y_3)}R(y_3,k)
\exp\Big(\frac{1}{2}\alpha'k^2\frac{(y_1-y_3)^2}{y_1-y_2}\Big).
\eeq

In the general case Eq. (\ref{rrand1}) can only be solved
numerically. In a particular case $a=1$ both the Green function
$G^{(0)}$ and $\Sigma_1$ depend only on the rapidity difference:
\beq p(y)=e^{\epsilon y} \eeq and \beq
G^{(0)}_f(y_1,y_2,k)=e^{-(y_1-y_2)(\epsilon+\alpha'k^2)}=G^{(0)}_f(y_1-y_2,k),
\label{gy} \eeq \beq
\Sigma_1(y_1,y_2,k)=-C\Big(\frac{1}{y_1-y_2}e^{-(y_1-y_2)(2\epsilon+\alpha'k^2/2)}
+\delta(y_1-y_2)\ln(c_Ry_{min}\Big)=\Sigma_1(y_1-y_2,k).
\label{sigmay} \eeq They have the same form as in the vacuum with
the opposite sign of $\epsilon$. So if one additionally neglects
the formally 3d-order contribution $\Sigma_2$ the Dyson equation
(\ref{grand}) can be again analytically solved by the Laplace
transform in the same way as in the vacuum case. Moreover, the
solution is exactly the same as in the vacuum case with the change
$\epsilon\to -\epsilon$: \beq G_f^{-1}(E,k)=E+\epsilon+\alpha'k^2-
C[\ln(E+2\epsilon+\alpha'k^2/2)+C_E-\ln c_R]. \eeq Singularities
of $G_f(E,k)$ in the complex $E-$ plane consist of a left cut
starting at $E_c=-2\epsilon-\alpha'k^2/2$ and corresponding to the
standard cut generated by a subcritical pomeron with intercept
$1-\epsilon$ and a possible pole whose position depends on the
renormalization constant $c_R$ and which determines the asymptotic
at large rapidities. With the choice $(\ref{cr0})$ corresponding
to $\Sigma_2=0$ we have \beq G_f^{-1}(E,k)=E+\epsilon+\alpha'k^2-
C\ln\frac{E+2\epsilon+\alpha'k^2/2}{2\epsilon} \eeq The Green
function vanishes at $y\to\infty$ as $e^{-y\epsilon_0}$ where for
$\alpha'k^2<2\epsilon$ \beq \epsilon+\alpha'k^2<\epsilon_0\leq
2\epsilon+\alpha'k^2/2. \eeq If $\alpha'k^2\geq 2\epsilon$, the
two-pomeron cut moves to the right of the pole, which splits into
two complex conjugate poles on the physical sheet, and the
asymptotic is determined by the cut so that
$y_0=2\epsilon+\alpha'k^2/2$ imaginary parts of the two terms in
the denominator are opposite.

Unfortunately inclusion of $\Sigma_2$ spoils invariance with respect to
translations in rapidity, and solution of the Dyson equation
by means of the Laplace transform becomes impossible.

\section{Numerical illustration}
In this section we report on numerical results for the amplitude
$T(y)$ and Green function $G_f(y,k)$ which follow from the Dyson
equations (\ref{trand}) and (\ref{grand}) with the pomeron
self-mass $\Sigma=\Sigma_1+\Sigma_2$ given by loops of Fig. 2 $a$
and $b$ and realistic values of $\epsilon$, $\alpha'$, $\lambda$
and $g$ For the latter we take the standard values, which
correspond to the experimental data at comparatively low energies
\beq \epsilon=0.08, \ \ \alpha'=0.2\ \ GeV^{-2},\ \ \lambda=-0.48
GeV^{-1},\ \ g=5.94\ \ GeV^{-1} \label{par} \eeq For the
transverse nuclear density we choose that in the center of the
nucleus with the constant nuclear density within a sphere of
radius $R_A=A^{1/3}\cdot 1.15$ fm.

With $\epsilon$, $\alpha'$ and $\lambda$ given by (\ref{par}) we obtain
from Eq. (\ref{reneps}) the renormalized values for the intercept and slope
in the vacuum:
\beq
\epsilon^{(r)}=0.155\pm i\,0.139,\ \ {\alpha'}^{(r)}=0.172\mp i\,0.043
\label{renin}
\eeq
Both are complex. The real part of the pomeron trajectory is found to
be linear in $k^2$ with a good precision. The imaginary part is not,
the value of Im ${\alpha'}^{(r)}$ diminishing from 0.043 at $k=0$
to 0.020 at $k=2$ GeV/c.  So we are dealing with a supercritical pomeron
with a complex intercept and slope.

Next we pass to the amplitude and pomeron Green function in the
nuclear background.

Fig. \ref{fig4} shows the ratio $r(y)$ of the calculated amplitude to the lowest order
one as a function of $y$ for two values of the atomic number $A=64$ and 207.
As one observes as rapidity grows the ratio initially goes up reaching at
its maximum values of the order 2.2 $\div$ 2.3 depending on $A$ and then
gradually goes to unity, as determined by the chosen value of $c_R$.
The $A$-dependence is in fact weak, so that the two curves are close
to one another. The rather large values of the ratio at intermediate
rapidities show that for the chosen parameters
the loop contribution is not at all small. In other words the
strength of the triple pomeron interaction  is relatively large although
it effectively goes down at large rapidities.

In Figs. \ref{fig5}-\ref{fig7} we show the  $R(y,k)$ of the Green
function to its lowest order value as a function of $y$ for values
of $k=0$, 1 and 2 GeV/c and $A=64$ and 207. For convenience for
$k=2$ GeV/c we plot $R_1(y,k)=-R(y,k)\exp\Big(y(\epsilon-\alpha'
k^2/2)\Big)$, since for such large $k$ the asymptotic is negative
and governed by the cut contribution. Again we observe that the
$A$-dependence is weak. The $y$-dependence strongly depends on the
chosen value of $k$, which is to be expected, since the cut
contribution vanishes at $k\to\infty$ much slower, with a twice
smaller slope. For all momenta the loop contribution is relatively
large at all rapidities, which again indicates that with our
choice $\lambda$ is not small. Oscillations observed in Figs.
\ref{fig5}
 and \ref{fig6} illustrate
that the 'physical' intercept $\epsilon^{(r)}$ for moderate $k$ in fact
splits into two complex conjugate values. At larger $k$ the asymptotic
is taken over by the cut contribution and oscillations disappear.

\section{Conclusions}
We studied the contributions of the two simplest loops in the local
reggeon field theory with a supercritical pomeron in the nucleus.
Our results show that the nuclear surrounding effectively transforms
the supercritical pomeron with the intercept $\alpha(0)-1=\epsilon>0$
into the subcritical one with the intercept $-\epsilon$. As a result,
at high energies the pomeron Green function vanishes and contributions
from multipomeron exchanges vanish still faster according to the standard
predictions for the subcritical pomeron.  With that, for a finite
nucleus with a constant profile function, the scattering amplitude
tends to a constant value at high energies. Contribution from
self-mass insertions do not change this behaviour, since in the rapidity
space they are dominated by the configuration in which these
insertions enter at small rapidities and the bulk of the
rapidity is covered by the lowest order amplitude. With an appropriate
choice of the renormalization constant one can make the contribution
from loops vanish at high energies. The renormalized pomeron
intercept in this theory (in vacuum) is complex
with a positive real part. This makes the conventional renormalization
technique inapplicable.

These results have been obtained for a simple case of a constant
nuclear profile function $T(b)$ and with only lowest order loops
taken into account. The limitation to a constant $T(b)$ is
technical and we believe that the same conclusions can be obtained
for a realistic $T(b)$. However this requires using non-trivial
pomeron Green functions in the external $b$-dependent field, which
can only be calculated numerically. Construction of loops in this
case presents a formidable calculational problem, which seems to
be outside our technical possibilities. The only difference we
expect with a realistic $T(b)$ is due to its long distance tail.
Then at  large $b>b_0(y)$ the amplitude and the Green function
will no more be damped by the nucleus and become purely
perturbative, that is growing as $e^{\epsilon y}$ at large $y$.
Very crudely we expect $b_0$ to be determined by the relation
$T(b_0)e^{\epsilon y}\sim 1$ which gives $b_0(y)\sim ay$ where $a$
is an $A$-independent parameter of the dimension of length
($a=0.545$ fm with the Woods-Saxon nucleus density). Then, again
very crudely, the cross-section obtained by the integration over
all $b$ will become proportional to $y^2$ in accordance with the
conclusions in ~\cite{ciaf}. Inclusion of more complicated loops
does not present any difficulty of principle either. For small
enough $\lambda$ their contribution will be small at all
rapidities and we do not expect any qualitative change.

Note that with the conventional values (\ref{par}) for $\epsilon$, $\alpha'$
and $\lambda$ the loop contribution proves to be not at all small,
in spite of the rather small value of the effective loop parameter
$C\sim 0.09$. The renormalized intercept (\ref{renin}) obtained as the
solution of Eq.(\ref{greenpo}) is quite different from its zero-order
value $\epsilon=0.08$, which testifies that the coupling constant $\lambda$
is in fact quite large. Because of that we cannot claim that
with the conventional parameters (\ref{par}) our results are
complete . More complicated
loops have to be included to make the results reliable and fit to be used
for the description of physical observables.

This is unfortunate, but we consider our results to be interesting mostly
from the purely theoretical point of view. They may serve as a basis
for treating loop contributions in the perturbative QCD. It seems to
be advantageous to study them in the nuclear surrounding, which makes
the high-energy behaviour much more tractable.

\renewcommand{\theequation}{\thesection\arabic{equation}}
\setcounter{equation}{0}
\appendix
\renewcommand{\thesection}{Appendix A: }
\section{Possible poles of  propagators }

\renewcommand{\thesection}{A}

Let the denominator in (\ref{gvac}) be \beq
D(E,k)=E-\epsilon+\alpha'k^2-
|C|[\ln(E-2\epsilon+\alpha'k^2/2)+C_E-\ln c_R], \label{denvac}
\eeq where we have taken into account that $C<0$. We denote
$\delta=|C|(C_E-\ln c_R)$ and rewrite (\ref{denvac}) as \beq
D(E,k)=E-\epsilon+\alpha'k^2-\delta-
|C|\ln(E-2\epsilon+\alpha'k^2/2).\eeq Function $D(E,k)$ has a left
cut starting at $E=2\epsilon-\alpha'k^2/2$, so that it can vanish
only to the right of the branchpoint. At $E\to
2\epsilon-\alpha'k^2/2$ from the right $D(E,k)\to +\infty$. At
$E\to +\infty$ also $D(E,k)\to +\infty$. The single minimum of
$D(E,k)$ as a function of $E$ occurs at \beq \frac{\partial
D(E,k)}{\partial E}=1-|C|\frac{1}{E-2\epsilon+\alpha'k^2/2}=0,
\eeq that is at \beq
E_{min}=e^{|C|}+2\epsilon-\frac{1}{2}\alpha'k^2. \eeq The minimal
value of $D(E,k)$ is \beq
D_{min}(E,k)=e^{|C|}+\epsilon+\frac{1}{2}\alpha' k^2-\delta-|C|^2.
\eeq Obviously if $D_{min}>0$ then $D(E,k)$ does not vanish on the
physical sheet of $E$. If $D_{min}<0$ there are two zeros to the
right of the cut. The rightmost zero $E_0$ determines the
asymptotic of the propagator at large $y$. The value of $D_{min}$
depends on the choice of $\delta$. So if \beq
\delta>e^{|C|}+\epsilon+\frac{1}{2}\alpha' k^2-|C|^2 \label{ineq}
\eeq the propagator develops two poles to the right of the
branchpoint at $E=2\epsilon-\alpha'k^2/2$, the larger of which is
located to the right of $E_{min}$ and thus to the right of the
unperturbed pole at $E=\mu-\alpha' k^2$. This larger pole takes
the function of the renormalized intercept. If, on the other hand,
relation (\ref{ineq}) is not satisfied, the propagator has no
poles on the physical sheet and its asymptotic at $y\to\infty$ is
determined by the branchpoint at $E=2\epsilon-\alpha' k^2/2$

Passing to the nuclear surrounding with $a=1$, at $k=0$ we find a
denominator \beq
D(E)=E+\epsilon-|C|\ln\frac{E+2\epsilon}{2\epsilon} \eeq We put
$E+2\epsilon=2\epsilon\xi$, $|C|/(2\epsilon)=\eta<<1$, so that
zeros are determined by the equation for $\xi$ \beq
f(\xi)=\xi-\eta\ln\xi-\frac{1}{2}=0. \eeq The minimum of $f(\xi)$
occurs at $\xi_{min}=\eta$ and \beq
f_{min}=\eta-\eta\ln\eta-\frac{1}{2} \eeq At $\eta<1$ $f_{min}<0$
for $\eta<0.186683$, for which values there exist two zeros \beq
0<\xi_1<\eta<\xi_2<\frac{1}{2} \eeq (the limiting value $1/2$ is
achieved by the larger zero at $\eta\to 0$). This means that poles
in the propagator exist for small enough $\eta$ and are located
between the unperturbed pole at $E=-\epsilon$ and branchpoint at
$E=2\epsilon$. One can easily see that this result is also true
for $k^2>0$. So the poles do not change the asymptotic of the
amplitude, but diminish the intercept of the propagator, making it
still more subcritical.

\renewcommand{\thesection}{Figure captions}
\section{}

\noindent Figure 1: The new vertex for two-pomeron annihilation,
which appears after the shift in field $\phid$.

\noindent Figure 2: Some simple loop diagrams for the pomeron
Green function.

\noindent Figure 3: Diagrams with one loop ($a,\, b$ and two loops
($c$ and $d$) for the scattering amplitude.

\noindent Figure 4: The ratio $r(y)$ of forward scattering
amplitude to its lowest order for Cu and Pb targets.

\noindent Figure 5: The ratio $R(y)$ of the pomeron Green function
to its lowest order at $k=0$ for Cu and Pb targets.

\noindent Figure 6: Same as Fig. \ref{fig5} at $k=1$ GeV/c.

\noindent Figure 7: The scaled ratio
$R_1(y,k)=-R(y,k)\exp\Big(y(\epsilon-\alpha' k^2/2)\Big)$ at $k=2$
GeV/c for Cu and Pb targets.

\newpage

\begin{figure}
\includegraphics[width=12 cm]{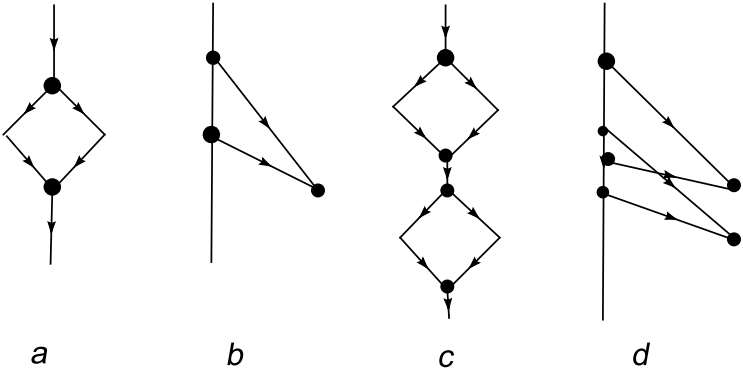} \caption{Some simple loop
diagrams for the pomeron Green function} \label{fig2}
\end{figure}

\begin{figure}
\hspace*{2 cm} \includegraphics[width=10 cm]{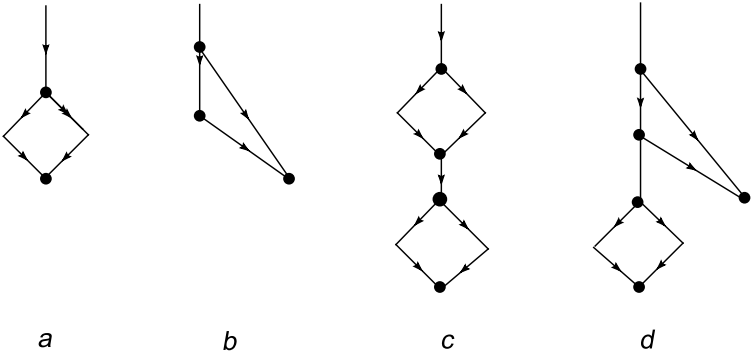}
\caption{Diagrams with one loop ($a,\, b$ and two loops ($c$ and
$d$) for the scattering amplitude} \label{fig3}
\end{figure}

\begin{figure}
\includegraphics[width=12 cm]{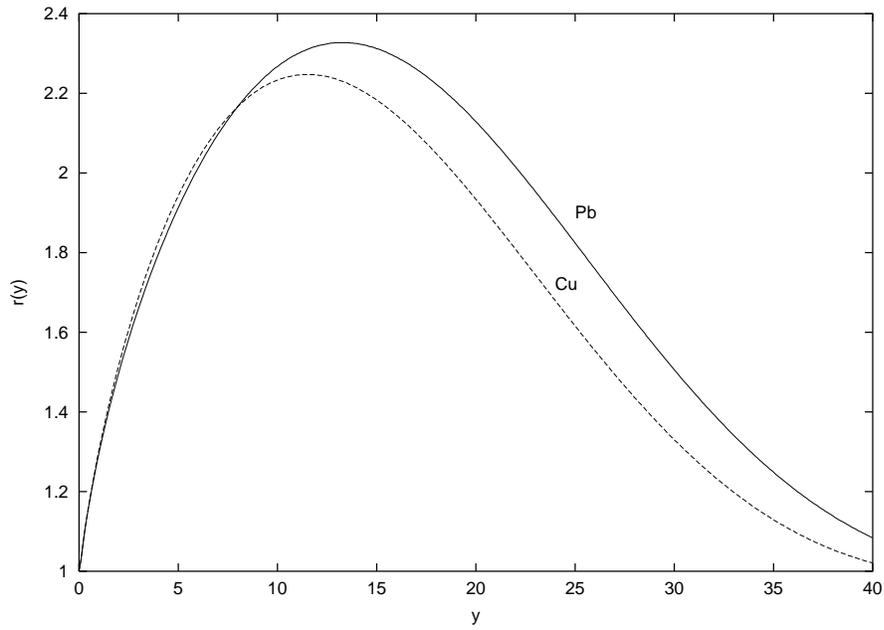} \caption{The ratio $r(y)$ of
forward scattering amplitude to its lowest order for Cu and Pb
targets} \label{fig4}
\end{figure}

\begin{figure}
\includegraphics[width=12 cm]{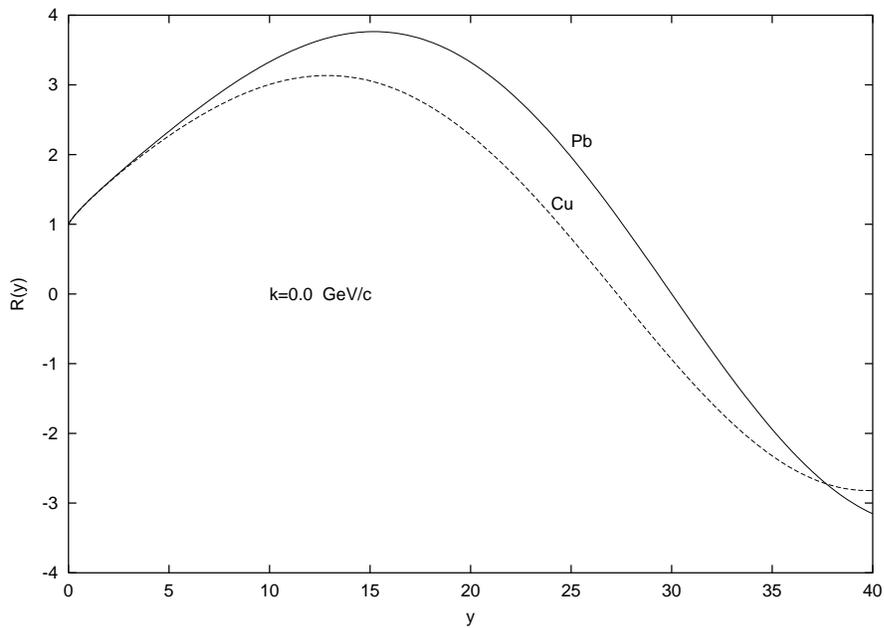} \caption{The ratio $R(y)$ of
the pomeron Green function to its lowest order at $k=0$ for Cu and
Pb targets} \label{fig5}
\end{figure}

\begin{figure}
\includegraphics[width=12 cm]{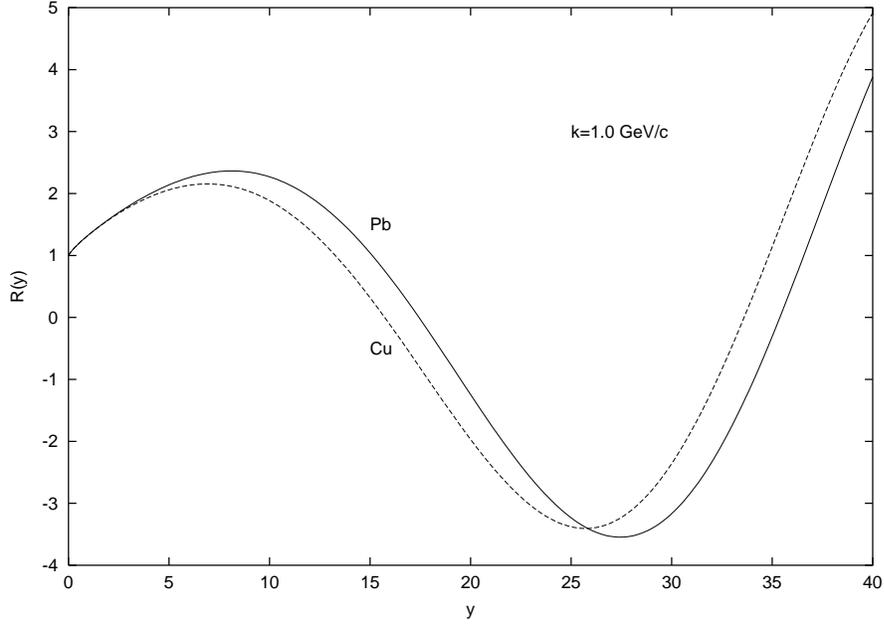} \caption{Same as Fig.
\ref{fig5} at $k=1$ GeV/c} \label{fig6}
\end{figure}

\begin{figure}
\includegraphics[width=12 cm]{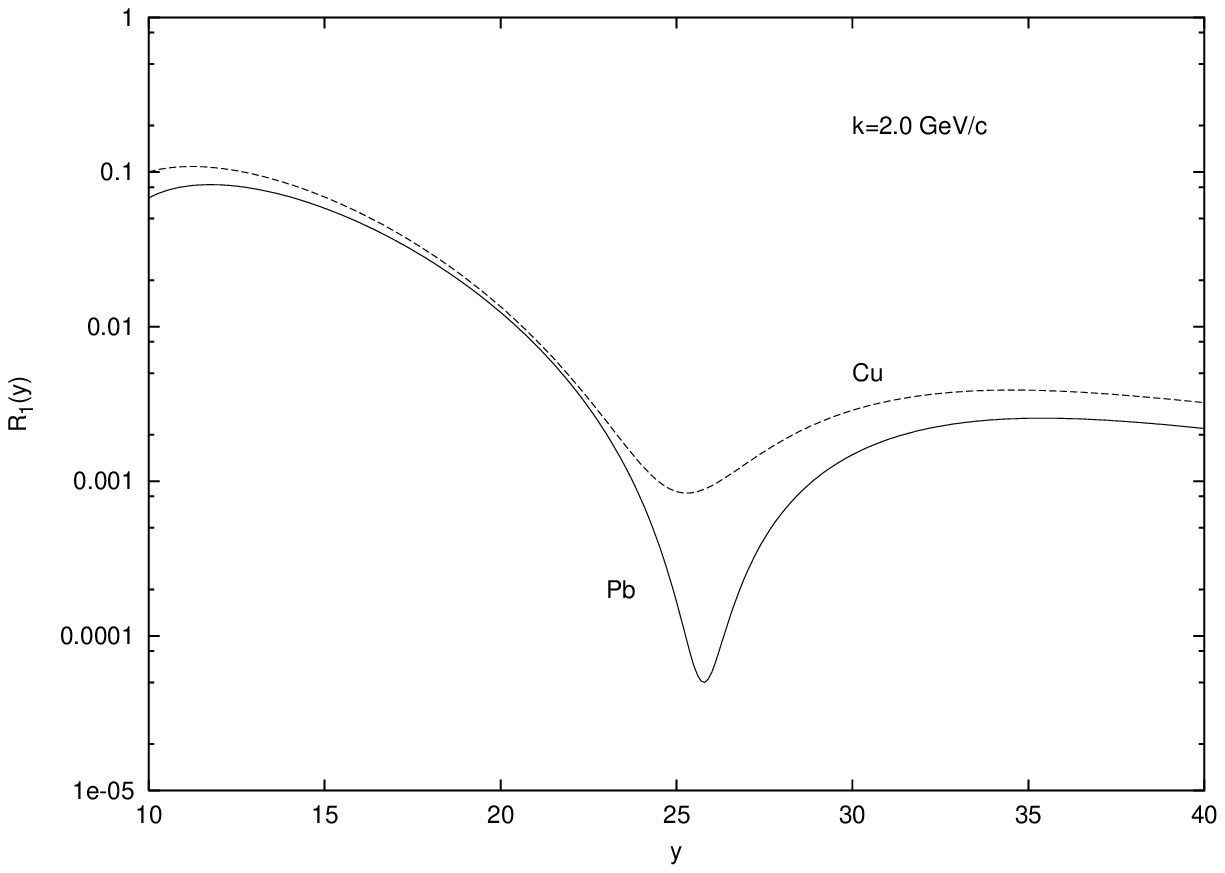} \caption{The scaled ratio
$R_1(y,k)=-R(y,k)\exp\Big(y(\epsilon-\alpha' k^2/2)\Big)$ at $k=2$
GeV/c for Cu and Pb targets} \label{fig7}
\end{figure}


\begin{thebibliography}{99}
%
\bibitem{bal} I.I.Balitsky, Nucl. Phys. B {\bf463}, 99 (1996)
%
\bibitem{kov} Yu.V.Kovchegov, Phys. Rev. D {\bf60}, 034008 (1999); {\bf61}, 074018 (2000)
%
\bibitem {bra1} M.A.Braun, Eur. Phys. J. C {\bf16}, 337 (2000)
%
\bibitem{schwim} A.Schwimmer, Nucl.Phys. B {\bf94}, 445 (1975)
%
\bibitem{amati} D.Amati, L.Caneschi, R.Jengo, Nucl. Phys. {\bf 101}, 397 (1975)
%
\bibitem{aless} V.Alessandrini, D.Amati, R.Jengo, Nucl.Phys. B {\bf108}, 425 (1976)
%
\bibitem{jengo} R.Jengo, Nucl. Phys. B {\bf108}, 425 (1976)
%
\bibitem{ciabel} M.Ciafaloni, M.Le Bellac, G.C.Rossi,
Nucl. Phys. B {\bf130}, 388 (1977)
%
\bibitem{bravac} M.A.Braun, G.P.Vacca, Eur. phys. J. C {\bf50}, 857 (2007)
%
\bibitem{abarb} H.D.Abarbanel, J.B.Bronzan, A.Schwimmer, R.Sugar,
Phys. Rev. D {\bf14}, 632 (1976)
%
\bibitem{ciaf}D.Amati, M.Le Bellac,
G.Marchesini, M.Ciafaloni, Nucl. Phys. B {\bf112}, 107 (1976)
%

\end{thebibliography}
\end{document}